\documentclass[a4paper]{aa}

\usepackage{graphicx}
\usepackage{txfonts}

\begin{document}

\title{A new radial system of dark globules in Monoceros}

   \subtitle{}

\author{C. Carrasco-Gonz\'alez\inst{1}
 \and R. L\'opez\inst{2}
 \and A. Gyulbudaghian\inst{3}
 \and G. Anglada\inst{1}
 \and C.W. Lee\inst{4}}

\institute{Instituto de Astrof\'\i sica de Andaluc\'\i a, CSIC, Camino Bajo de Hu\'etor 50, E-18008 Granada, Spain; e-mail: charly@iaa.es, guillem@iaa.es
 \and Departament d'Astronomia i Meteorologia, Universitat de Barcelona, Av. Diagonal 647, E-08028 Barcelona, Spain; e-mail: rosario@am.ub.es
 \and Byurakan Astrophysical Observatory, Aragatsotn prov. 378433, Armenia; e-mail: agyulb@bao.sci.am
 \and Korea Astronomy \& Space Science Institute, 61-1 Hwaam-dong, Yusung-gu, Taejeon 305-348, Korea; e-mail: cwl@kasi.re.kr}

\offprints{C. Carrasco-Gonz\'alez, \email{charly@iaa.es}}

   \date{Received October 5, 2005 / Accepted November 23, 2005}

  \abstract{}{We analyze the LBN 978 HII region in order to study the HII/molecular cloud interaction.}{We used the IAC-80 telescope to image the region with narrow-band filters and the Very Large Array to obtain 
   a radio continuum map at 3.6 cm. We also used the DSS2 red images and the NRAO VLA Sky Survey at 20 cm.}{We have discovered a new radial system of dark 
   globules associated with the LBN 978 HII region, containing a
   group of at least eight cometary bright-rimmed globules with the same morphological type. The brightest source is also detected in the radio continuum. Analysis of optical and radio
   emission suggest that this object is photoionized by HD 47432, the central star of the LBN 978 HII region.}{}
   
   
   \maketitle

\section{Introduction}

Several HII regions, driven by O and/or early B type stars, are associated with radial systems of dark globules. These molecular
globules appear in groups that are spread along the HII region surrounding the ionizing star, and with their axes of symmetry 
oriented towards the central star. Detailed studies of radial systems have been performed in the Rosette Nebula (Herbig 1974) and in 
the Gum Nebula (Reipurth 1983). In a paper by Gyulbudaghian et al. (1994) a list of 23 of such systems was reported. 

These dark globules are dense, cold and neutral molecular clouds. Some of them are also named bright-rimmed clouds (BRCs) because 
they present bright rims that trace the interphase between the ionized and the neutral fractions of the cloud. Systems of BRCs 
globules around early type stars were mentioned already by Pottasch (1956). Sugitani et al. (1991) and De Vries et al. (2002) enumerate 
three different morphological types of BRCs and suggest that they may actually represent a time evolution sequence. Globules with a 
type-C morphology, according this classification, are also named cometary globules. They present curved bright rims and bright tails, with 
the apex oriented towards the central photoionizing source of the HII region.

The investigation of radial systems of dark globules is important for evolution problems of molecular clouds and the effects of the UV
radiation from central stars. BRCs offer a good opportunity for studying the various phenomenon that occur in HII region-molecular 
cloud interactions and to investigate the phenomenon of triggered star formation in such globules due to the compression by ionization
front shocks. Evidence for star formation in some globules has been revealed by studies of IRAS point-source association, from 
observations of outflows in the globules and by the detection of H$_2$O maser emission (see Patel et al. 1995, Valdettaro et al. 2005, and 
references therein). 

In this Letter, we present an analysis of DSS2 red images and radio continuum data from the NRAO VLA Sky Survey (NVSS; Condon et al.
1998) of the HII region LBN~978, revealing a system of bright rims facing to the photoionizing star, suggesting the presence of a new 
radial system of dark globules associated with this HII region. We report narrow-band filter optical observations of a $\sim7'\times7'$ area 
around the compact nebula CN3 (Gyulbudaghian et al. 1990) as well as 3.6 cm VLA observations, that allow us to identify a group of bright 
rimmed cometary globules, confirming the existence of the proposed radial system of dark globules in the LBN 978 HII region. Finally, we 
discuss on the nature of CN3.

\section{Observations}

Deep CCD narrow-band optical images of the CN3 field were obtained with the 0.82~m IAC-80 telescope of the Observatorio del Teide 
(Tenerife, Spain) in several observing runs, from November 2004 to April 2005 using the Service Time mode facility. Observing conditions 
were not photometric (typical values of the seeing were 1$\farcs$8). The images were obtained with the IAC-80 CCD camera (image scale, 
0$\farcs$4325~pixel$^{-1}$; sampled field, $7\farcm$3$\times$7$\farcm$3) trough five narrow-band filters (three line-emission filters 
and two more filters centered on the continua nearby the lines, being the ``continuum'' filters useful to discriminate between scattered 
starlight and pure-line emission (Table \ref{tabla_rosario}). Astrometric calibration of the images was performed using $\alpha$, 
$\delta$ (J2000) coordinates of thirteen field stars obtained from the USNO-B1.0 Catalogue\footnote{The USNOFS Image and Catalogue 
Archive is operated by the United States Naval Observatory, Flagstaff Station}). The accuracy of the calibration gives a rms $\leq$0$\farcs$3 
in both coordinates.

\begin{table}
\caption{Optical Observations}
\label{tabla_rosario}
\centering
\begin{tabular}{cccc}
\hline \hline
	       & Central	 &	 &	    \\ 
	      & wavelength	& FWHM  & $t$$_{\rm exp}$\\ 
  Filter       &    (\AA)	 & (\AA) & (s)      \\ \hline
 H$\alpha$ (l) & 6571		 &  47   & 2700     \\
 H$\alpha$ (c) & 6687		 &  50   & 2700     \\
 $[$SII$]$ (l) & 6724		 &  50   & 1800     \\
 $[$SII$]$ (c) & 6803		 &  50   & 2700     \\
 $[$NII$]$ (l) & 6589		 &   9   & 2700     \\ \hline
\end{tabular}
\end{table}

The radio continuum observations were carried out on 2004 March 27 at 3.6 cm using the Very Large Array (VLA) of the National Radio
Astronomy Observatory (NRAO\footnote{NRAO is a facility of the National Science Foundation operated under cooperative agreement by
Associated Universities, Inc.}) in the C configuration. A 
tapering of 30~k$\lambda$ was used in order to emphasize the extended emission. The resulting synthesized beam size was 
6$\farcs$4$\times$5$\farcs$6 with P.A. = $-$28$\degr$, and we achieved a rms noise of 15 $\mu$Jy beam$^{-1}$.

\section{Results}

\subsection{New radial system}

\begin{figure}
\resizebox{\hsize}{!}{\includegraphics{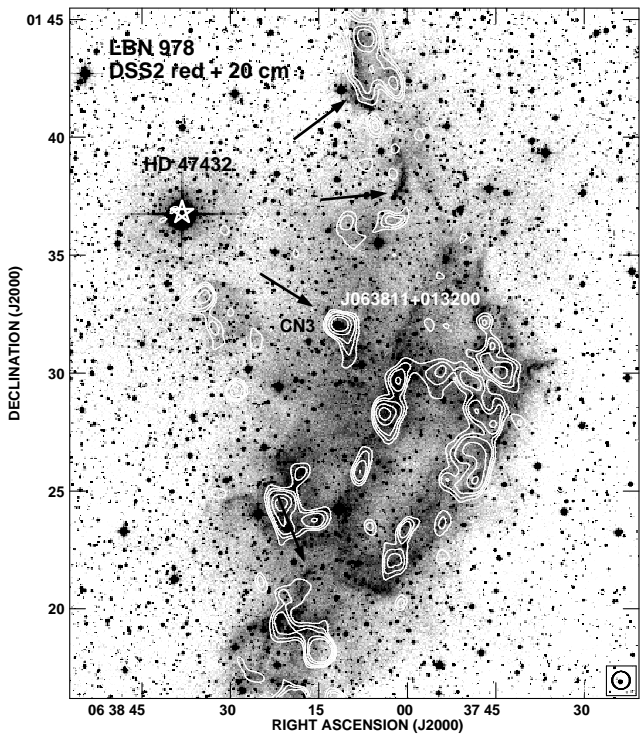}}
\caption{DSS2 red image of the HII region LBN 978 overlaid with NVSS contours at 20 cm. Contour levels are $-$3, 3, 4, 5, 7
and 9 times the rms of the map, 300 $\mu$Jy beam$^{-1}$. The star symbol marks the position of HD
47432 (SAO 114191), the ionizing source of LBN 978. Some of the bright rims are indicated by arrows. The source NVSS J063811+013200 
coincides with the bright nebulosity CN3.}
\label{fig1}
\end{figure}

In Fig. \ref{fig1} we show the DSS2 red image of the LBN 978 HII region overlaid with the NVSS radio continuum contours at 20~cm. 
In the DSS2 image, the LBN 978 HII region appears as a curved nebula surrounding the star HD~47432 (SAO 114191), which 
has been proposed as its ionizing source (Sharpless 1959; Felli \& Harten 1981). Several bright rims facing to the star can be 
identified in this image. The NVSS at 20~cm reveals knotty emission covering an extended region of $\sim30'$ that appears to follow 
the brightest rims of the optical emission of LBN~978 (see Fig. \ref{fig1}). The spatial distribution of the rims, surrounding the star HD~47432 at a fairly uniform projected distance, led us to propose that they contain a new radial system 
of dark globules. 

From the V magnitude, the (B-V) color given by Johnson et al. (1966), and spectral type O9.5II (Kharchenko 2001) of the star 
HD~47432, and adopting the usual value R$_V$=3.2, we derive a distance of $\sim$1250 pc (which is similar to the estimates given 
by Felli \& Harten 1981, and Blitz et al. 1982). The star HD~47432 is located near the OB-association Mon OB 2, which is at a distance 
of 1550 pc (Ruprecht et al. 1970), so it is probable that HD~47432 is a member of the Mon OB 2 association.

As can be seen in Fig. \ref{fig1}, the source NVSS J063811+013200 appears to coincide with CN3, a bright optical compact nebula reported by
Gyulbudaghian et al. (1990). As observed by the NVSS (with an angular resolution of $\sim40''$) the 20 cm flux density of this source 
is $4.6\pm0.6$ mJy and the source appears elongated with a size of $\sim1'$. 

\subsection{New group of cometary bright-rimmed globules}

\begin{figure}
\resizebox{\hsize}{!}{\includegraphics{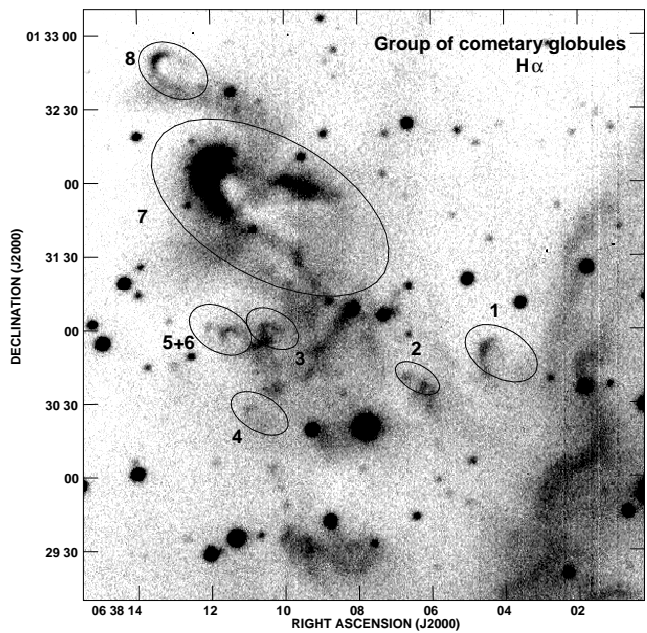}}
\caption{H$\alpha$ image ($t_{\rm exp}$=2700 s) of the new group of bright-rimmed cometary globules. The brightest rim corresponds to CN3.}
\label{fig2}
\end{figure}

In Fig. \ref{fig2} we show a $\sim 4'\times 4'$ field of our H$\alpha$ image, corresponding to the region immediately surrounding CN3, 
located about $\sim8'$ SW of the star HD~47432 
(the ionizing star of the HII region). The image reveals a group of features 
presenting a similar morphological structure, with an arc-shaped bright rim facing the star HD~47432 and a long body of dark globule at 
P.A.$\simeq$ 60$\degr$, oriented towards the central star. We identify at least eight of such cometary globules in this area, and in 
Table \ref{tabla_armen} we list the central positions of their rims. 

\begin{table}
\caption{Cometary bright-rimmed globules$^{\rm a}$.}
\label{tabla_armen}
\centering
\begin{tabular}{ccc}
\hline \hline
Number   & $\alpha$(J2000)$^\mathrm{b}$ & $\delta$(J2000)$^\mathrm{b}$ \\ \hline
1        & 06 38 04.4      &  +01 30  55 \\
2        & 06 38 06.6      &  +01 30  44 \\
3        & 06 38 10.7      &  +01 31  03 \\
4        & 06 38 11.1      &  +01 30  31 \\
5        & 06 38 11.8      &  +01 31  01 \\
6        & 06 38 12.3      &  +01 31  03 \\
7$^\mathrm{c}$    & 06 38 12.4    &  +01 31  07 \\
8        & 06 38 13.7     &  +01 32  49 \\ \hline
\end{tabular}
\begin{list}{}{}
\item[$^\mathrm{a}$]Positions listed correspond approximately to the apex of the bright rims. Uncertainty is $\sim$1$\arcsec$.
\item[$^\mathrm{b}$]Units of right ascension are hours, minutes and seconds. Units of declination are degrees, arcmins and arcsecs.
\item[$^\mathrm{c}$]Corresponds to CN3.
\end{list}
\end{table}

The [SII] and [NII] emissions are weaker than the H$\alpha$ emission for all of the globules detected. We believe that all these
features are not reflection nebulae of embedded stars associated with the globules, because we have not detected appreciable emission at 
the positions of these nebulae in our ''continuum'' images.

Therefore we propose that these features constitute a group of cometary BRCs, where the emission from a photionization front and 
shock-excited HH-like objects may coexist (see, eg, Ogura et al. 2002). 

\subsection{The CN3 cometary globule}

The brightest source in the field is CN3, the bright rim of globule 7 (see Fig. \ref{fig2} and Table \ref{tabla_armen}). This object shows a
well defined arc-shaped rim, and a dark globule with a size of $\sim30''$, corresponding to $\sim38000$ AU. 

\begin{figure}
\resizebox{\hsize}{!}{\includegraphics{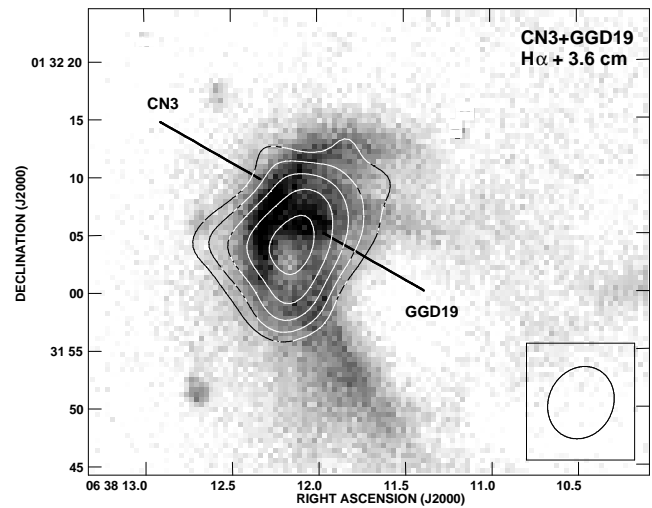}}
\caption{H$\alpha$ image ($t_{\rm exp}$=900 s) of CN3 overlaid with the 3.6~cm continuum contour map. Contour levels are $-$3, 3, 5, 8, 12 
and 17 times the rms of the map, 15 $\mu$Jy beam$^{-1}$.}
\label{fig3}
\end{figure}

CN3 is the only source in the field detected in our 3.6 cm VLA observations, with a flux density of 0.78$\pm$0.02 mJy. The other features are
expected to be weaker than CN3 in the radio continuum, since they are also weaker in H$\alpha$. The 3.6 cm source is 
resolved by the $\sim5''$ beam of the VLA, and the shape of the emission follows the optical rim, but the position of the peak 
($\alpha$(J2000)=06$^h$ 38$^m$ 12$\fs$15, $\delta$(J2000)=01$\degr$ 32$\arcmin$04$\farcs$1 $\pm$ 0$\farcs$2), as obtained 
from a Gaussian fit, is displaced $\sim5''$ SW from the apex of the optical rim (Table \ref{tabla_armen}). We note that there is a compact
emission peak in the H$\alpha$ image (see Fig. \ref{fig3}) that we tentatively identify with the HH-like object GGD 19 (Gyulbudaghian et
al. 1978\footnote{We note that there is a typographical error in the right ascension of GGD 19 as given in the paper by Gyulbudaghian et 
al. (1978).}); this object is equivalent to GM 2-37 of Gyulbudaghian \& Maghakian 1977. The deconvolved size of the 3.6 cm source is 
$\theta\simeq$8$\arcsec\times$4$\arcsec$ with P.A.=150$\degr$. Since the source is extended, and given the large difference in the 20 cm 
and 3.6 cm beam sizes, we cannot derive a reliable spectral index. 

The [NII]/H$\alpha$ and [SII]/H$\alpha$ line ratios in CN3 are indicative of the excitation of the object spectrum. Unfortunately, 
because all the optical runs were made under non-photometric conditions, images were not flux-calibrated. Thus, we cannot perform 
an accurate enough evaluation of the [NII]/H$\alpha$ and [SII]/H$\alpha$ line ratios in CN3. However, we can use our three 
narrow-band line images (after appropriate scaling using field stars close to the object, and substraction of the [NII] from the 
H$\alpha$+[NII] emissions) to perform a rough estimate of the relative strength of the H$\alpha$, [NII] and [SII] emissions in CN3. Low 
values of the [SII]/H$\alpha$ line ratio (0.2-0.5) are typically found in photoionized regions 
(e.g., Cant\'o 1981, McCall et al. 1985). We found [NII]/H$\alpha\simeq$ 0.2 and [SII]/H$\alpha\simeq$ 0.4 
(with an estimated error of $\la$30\%) around the peak emission of CN3. Low values of the [SII]/H$\alpha$ and [NII]/H$\alpha$ 
ratios are also found in the other globules in the field. Although these ratios only give us a rough indication of the relative strength 
of the line emissions and they must be taken with caution, we note that the derived values are fully in agreement with those obtained in 
photoionized regions. 

Free-free emission at centimeter wavelengths has been detected in several other BRCs (e.g. Baars \& Wendker 1976; 
Matthews 1979; Schwartz 1985) and can be used to estimate the ionizing photon rate required to account for the observed free-free
emission, providing a way to test the photionization vs.\ shock-excited hypotheses. Assuming that the emission from the 3.6 cm source 
associated with CN3 is optically thin, the ionizing photon flux required to maintain the ionization in this source can be estimated from 
eq. (6) in Lefloch et al. (1997),
$$
\Phi=1.24\times 10^{10} \left[ \frac{S_\nu}{\rm mJy} \right] \left[ \frac{T_e}{\rm K} \right]^{0.35} \left[ \frac{\nu}{\rm GHz}
\right]^{0.1}\left[ \frac{\theta_1\times\theta_2}{\rm arcsec^2} \right]^{-1}\rm cm^{-2}~s^{-1}.
$$
From the flux density (S$_\nu$=0.78 mJy) and the size ($\theta_1\times\theta_2=8''\times4''$) of the 3.6~cm source, and assuming 
an electron temperature $T_e$=10$^4$~K, we estimate that an ionizing photon flux of $\Phi\simeq9\times10^9$~cm$^{-2}$~s$^{-1}$, at 
the source position, is required. The spectral type of HD~47432 implies an ionizing photon flux of 
$\Phi_*=4.6\times10^{23}$~cm$^{-2}$~s$^{-1}$ at the stellar radius R$_*$= 14.7~R$_{\sun}$ (Vacca et al. 1996). Given that the separation 
between the star and CN3 is $\sim8'$, we estimate that the stellar ionizing photon flux of HD~47432 at the position of CN3 is
$\Phi=~6\times10^{9}$~cm$^{-2}$~s$^{-1}$, which is similar to the value required to produce the observed radio continuum flux. 

Therefore, the orientation of the apex of CN3 to the star, and the agreement in the required flux of ionizing
photons from the star suggest that CN3 is most likely a BRC with a cometary morphology. Sensitive higher angular resolution radio 
continuum observations could show the details of the ionized structure of this source, and reveal the presence of HH-like objects and/or 
embedded young stellar objects in this globule, as well as help to detect the remaining weaker globules. 

\section{Conclusions}

Using the DSS2 red images and NVSS radio maps, we have identified a new radial system of dark globules in the LBN 978 HII region,
surrounding the star HD~47432.  

Our narrow-band filter images have revealed a system of at least eight cometary globules with bright rims, of which CN3 appears as the 
brightest one. All of this globules are elongated, with symmetry axes oriented towards the star HD~47432, and present arc-shaped rims 
facing this star. The [NII]/H$\alpha$ and [SII]/H$\alpha$ line ratios and the lack of nearby continuum counterparts are fully in 
agreement with those expected for photoionized regions. We have detected 3.6 cm continuum emission associated with CN3, with a
flux density consistent with photoionization by the star HD~47432. 

It is interesting to note that all these BRCs are located approximately at the same distance from the ionizing star, suggesting a simultaneous 
formation of all of them. This fact, together with the fact that all of these BRCs show the same morphological type, give support to the 
time-evolution sequence of the morphological type of the BRCs proposed by Sugitani et al. (1991).

This system is an interesting target for future, more sensitive, 
optical/IR/radio studies. Higher angular resolution observations would help to understand the detailed structure of the ionization, and 
optical spectroscopy of higher excitation lines with larger telescopes would be helpful to search for 
radiative shock emission (HH objects and jets) similar to what it has been found in other bright rimmed clouds associated with HII regions.

\emph{Acknowledgements}. We acknowledge the Support Astronomer Team of the IAC for obtaining the optical images. We thank an anonymous referee for 
useful comments. G. A. acknowledges partial 
support from grant AYA 2002-00376 (Spain), and from Junta de Andaluc\'{\i}a (Spain). R.L. acknowledges support from grant AYA 2002-00205. 
C.W.L. acknowledges supports from the Basic Research Program (KOSEF R01-2003-000-10513-0) of the Korea Science and Engineering Foundation.

\end{document}